\documentclass[letter,twocolumn]{revtex4}
\usepackage{graphicx}
\usepackage{subfigure}
\usepackage{xcolor}
\usepackage{soul}
 \expandafter\let\csname equation*\endcsname\relax 
 \expandafter\let\csname endequation*\endcsname\relax 
\usepackage{amsmath}
\usepackage{wasysym}
\usepackage{amssymb}
\usepackage{color}
\usepackage{soul}
\usepackage{ulem}

\begin{document}
\title[Preferred degree networks]{Modeling interacting dynamic networks: III. Extraordinary properties in a
population of extreme introverts and extroverts}

\author{Wenjia~Liu$^{1}$, Florian Greil$^{2,3}$, 
K.\,E.\,Bassler$^{2,3,4}$, B.\,Schmittmann$^{1}$, 
and R.\,K.\,P.\,Zia$^{1,4,5}$}

\affiliation{$^1$Department of Physics and Astronomy, Iowa State University,
Ames, IA 50011, USA} 
\affiliation{$^2$Department of Physics, Department of
Physics, University of Houston, Houston, TX 77204, USA} 
\affiliation{$^3$Texas
Center for Superconductivity, University of Houston, Houston, TX 77204, USA} 
\affiliation{$^4$Max-Planck-Institut f\"{u}r Physik komplexer Systeme, N\"{o}thnitzer
Stra.\,38, D-01187~Dresden, Germany} 
\affiliation{$^5$Department of Physics,
Virginia Polytechnic Institute and State University, Blacksburg, VA 24061,
USA}
\author{E-mail: {\tt wjliu@iastate.edu, fgreil@uh.edu, bassler@uh.edu,
schmittb@iastate.edu, rkpzia@vt.edu}}
\begin{abstract}
Recently, we introduced dynamic networks with preferred degrees, showing
that interesting properties are present in a single, homogeneous system as
well as one with two interacting networks \cite{LiuJoladSchZia13,LiuSchZia14}. 
While simulations are readily performed, analytic studies are challenging,
due mainly to the lack of detailed balance in the dynamics. Here, we
consider the two-community case in a special limit: a system of extreme
introverts and extroverts - the $XIE$ model. Surprising phenomena appear,
even in this minimal model, where the only control parameters are the
numbers of each subgroup: $N_{I,E}$. Specifically, an extraordinary
transition emerges when $N_{I}$ crosses $N_{E}$. For example, the fraction
of total number of $I$-$E$ links jumps from $\thicksim 0$ to $\thicksim 1$.
In a $N_{I}=N_{E}$ system, this fraction performs a pure random walk so that
its distribution displays a \textit{flat plateau} across most of 
$\left[ 0,1\right] $, with the edges vanishing as $\left( N_{I,E}\right) ^{-0.38}$
for large systems. Thus, we believe the $XIE$ model exhibits an `extreme'
Thouless effect \cite{BarMukamel14}. 
For this limiting model, we show that detailed balance is restored and 
explicitly find the microscopic steady-state distribution. We then use a 
mean-field approach to find analytic expressions for the degree distributions 
that are in reasonably good agreement with simulations, provided $N_{I}$ is not
too close to $N_{E}$. 
\end{abstract}

\maketitle

\section{Introduction}

For many years, there has been much interest in characterizing and
understanding complex networks in nature \cite%
{Strogatz01,AlbertBarabasi02,DorogovtsevMendes02,Newman03,EstradaFoxHighamOppo10}%
. While much of the venerable network literature is devoted to static
aspects (e.g., graph theory), interest in their dynamic aspects began to
emerge recently \cite{BarratBarthelemyVespignani08,DorogovtsevGAMJ08},
especially in the context of adaptive co-evolutionary networks \cite%
{GrossDCBB06,GrossBlasius08}. Now, in the settings of both artificial
structures and natural environments, networks of very different kinds are
intimately entangled, e.g., the internet and transportation networks, ocean
currents and marine food-webs, etc. Thus, understanding how diverse,
interdependent networks interact is also important. While some work along
such lines \cite%
{RinaldiPeerenboomKelly01,PanzieriSetola08,BuldyrevParshaniPaulStanleyHavlin10,Vespignani10,BuldyrevShereCwilich11,KurantThiran06}
exists, the overall picture is far from clear. Within this context, our goal
is modest: What are the effects of coupling \textit{dynamic }networks of a
similar type, but with different characteristics?

To focus our effort, we limit ourselves to a special class of dynamic
networks, in which a node can cut or add links to the others. Various dynamics of this sort
are possible. (See, for example, \cite{LiuBassler06,DelGenioGross11}.) Here, though, nodes alter their links in an attempt to achieve a `preferred' number ($\kappa $) of connections. We believe that
the notion of preferred degrees is particularly suitable for modeling
networks in a social setting, in which introverts/extroverts are naturally
inclined to have few/many contacts. While such a model may be applicable to
the dynamics of opinion formation, it is also ideal for studying the effects
of self-imposed or public policies, in the event of an epidemic outbreak, by
including a time dependent $\kappa $. Thus, an individual's preferred degree
may drop by an order of magnitude or more, e.g., from hundreds of contacts
during a typical day, to just a few family members or care-givers.

Introduced a few years ago \cite{PZ2010,CSP24,CSP25,JoladLiuSchmittmannZia12}, 
we recently began more
systematic studies on interacting networks with preferred degrees, the
present being the third of a series. In the first \cite{LiuJoladSchZia13},
we provided specifics of how to implement a preferred degree for a
homogeneous population of individuals, introducing various natural ways to
incorporate the action of cutting/adding a link. With simple arguments, we
can predict the degree distribution, $\rho \left( k\right) $, which agrees
well with simulation results. To model interactions, we considered two such
communities, with different numbers ($N_{1,2}$) and $\kappa $'s,
and introduced a simple way to couple them: When a node attempts to add/cut
a link, it chooses a partner from the other community with probability $\chi 
$ (and an intra-community partner with $1-\chi $) for the action. Though
such a model seems minimal, the complexities involved include a
six-dimensional parameter space ($N_{1,2},\kappa _{1,2},\chi _{1,2}$) and
the presence of four degree distributions of interest (associated with
intra- and inter-community links). Despite its simplistic appearance, this
model displays some surprising phenomena. In the second of the series 
\cite{LiuSchZia14}, we performed a more detailed study in a wider region
of parameter space and arrived at some general principles to predict the
observed behavior. Nevertheless, since the dynamics does not obey detailed
balance, determining the microscopic distribution in the stationary state 
($\mathcal{P}^{ss}$, \`{a} la the Boltzmann distribution for systems in 
equilibrium) is a major challenge, let alone computing averages of 
macroscopic observables. In this paper, we consider a special limit of 
a system with two communities. Dubbed the \textit{XIE }model, it consists of 
a collection of e\textit{X}treme \textit{I}ntroverts and \textit{E}xtroverts,
whose only action is
either cutting a link or making one. This simplifying assumption means the
model has only two control parameters, $N_{I,E}$, the number of introverts
and extroverts. As the $N_{I}=N_{E}$ line is crossed, a remarkably sharp
transition emerges. It turns out that detailed balance is restored in this
limit and so, we are able to arrive at an explicit expression for 
$\mathcal{P}^{ss}$. Though computing averages with $\mathcal{P}^{ss}$ is still
difficult, various mean-field techniques can be applied and some aspects of
this model can be well understood. In particular, the transition appears to
be of mixed order and is likely to be in the class of an `extreme' Thouless
effect \cite{BarMukamel14}.While preliminary results have been reported
earlier \cite{CSP25,LiuSchmittmannZia12}, we will present a more systematic
study of this model here.

The rest of this paper is organized as follows. In the next section, we
provide a detailed description of the model, the master equation associated
with the stochastic process, and the exact microscopic distribution in the
steady state. The following section is devoted to studies of the degree
distributions, through both Monte Carlo simulations and a self-consistent
mean-field theory. In simulations with $N_{I}+N_{E}$ fixed at $200$ and
varying $N_{I}-N_{E}$, we find good agreement with the theory \textit{except}
the $N_{I}=N_{E}=100$ case. This surprising result leads us to explore, in
Section $4$, the behavior of a genuinely macroscopic quantity, $X$, the total
number of cross-links. We end with a summary and outlook.

\section{A model of extreme introverts and extroverts}

In our previous studies \cite{LiuJoladSchZia13,LiuSchZia14}, we introduced
adaptive networks in which each node adds/cuts links to other nodes
according to its `preferred degree' ($\kappa $) for both a homogeneous
population and a heterogeneous system of two communities with different
characteristics (e.g., numbers of nodes and $\kappa $'s). For the reader's
convenience, we summarize briefly here the specifications and the findings
for the simplest model, the extreme limit of which is the focus of this
study.

First, we consider a homogeneous population consisting of a dynamic set of
links among $N$ nodes, all of which are assigned the same $\kappa $.
In each time step, a random node is chosen and its degree, $k$, is noted. If 
$k>\kappa $, the node cuts one of its existing links at random. Otherwise,
it adds a link to a randomly chosen node not connected to it. (Models with
more `flexibility' in the nodes were studied in \cite{LiuJoladSchZia13}.)
This simple stochastic process is ergodic, so that the system eventually
settles into a unique stationary distribution, $\mathcal{P}^{ss}\left( 
\mathbb{A}\right) $ over the space of networks -- symmetric $N\times N$
adjacency matrices, $\mathbb{A}$. Despite the apparent randomness, the
resulting degree distribution is a Laplacian, $\rho (k)\propto \exp \left[
-\left\vert k-\kappa \right\vert \ln 3\right] $, which differs considerably
from the binomial distribution (i.e., Gaussian or Poisson, in the limit of 
large $N$) in Erd\H{o}s-R\'{e}nyi graphs \cite{E-R}.
Second, since our goal is interaction of networks, we turn to a study of two
networks of preferred degrees. Assuming that $\kappa _{1}<\kappa _{2}$,
naturally we refer to the first group as `introverts' ($I$) and the later one
as `extroverts' ($E$). To model the interaction between these groups, we
introduced a simple parameter $\chi \in \lbrack 0,1]$, which is the
probability for the chosen node to take action (adding or cutting) on a
cross-link. Letting $\chi _{1}\neq \chi _{2}$ as well as $N_{1}\neq N_{2}$,
we find a variety of behaviors \cite{LiuSchZia14}, some of which can be
understood, at least qualitatively. The most surprising phenomenon observed
concerns the statistical properties of the number of cross-links, $X$.
Specifically, we simulated two communities with identical properties: $%
N_{1,2}=100,\kappa _{1,2}=25,\chi _{1,2}=0.5$. In the stationary state, the
distribution of $X$ (dropping superscript $ss$ for simplicity and denoted by 
$P\left( X\right) $) displays a very broad and flat plateau. By contrast, we
can arbitrarily partition a \textit{homogeneous} population of $N=200$ into
two equal halves. Using $\kappa =25$ and defining $X$ as the number of links between
the halves, we find that $P(X)$ is just a binomial distribution, with
easily predictable parameters. As the standard deviation of the former is an
order of magnitude larger than the binomial, it is clear that, despite its
conceptual simplicity, coupling two networks via $\chi $ has a profound
effect on the system.

\begin{figure}[tbp]
\centering
\includegraphics[width=3.5in]{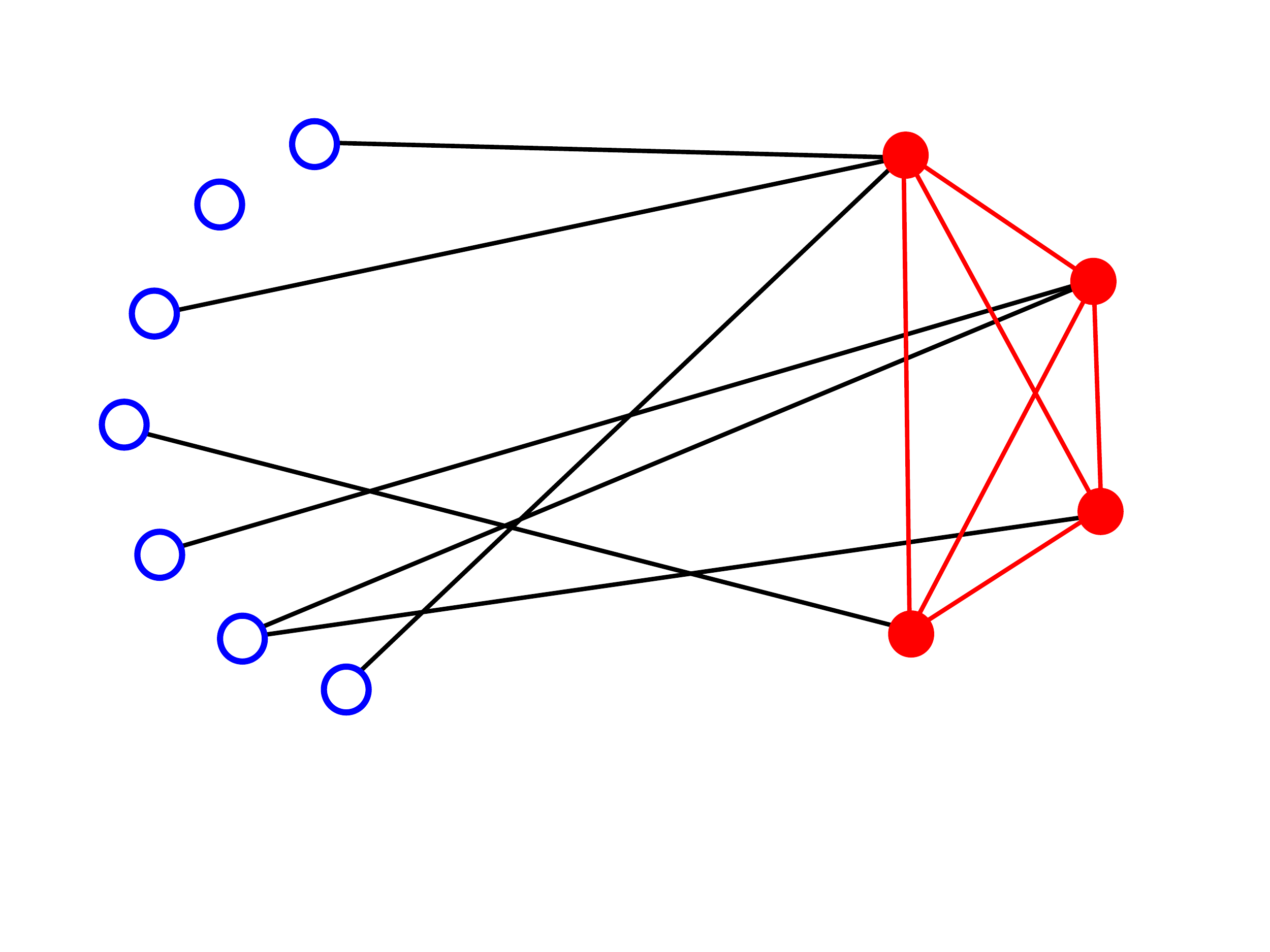}\vspace{-1.8cm}
\caption{The nodes of the two groups are denoted by circles: blue open ($I$)
and red closed ($E$). The black lines represent the active cross-links and
the red lines, the frozen $E$-$E$ links. For this network, the sets of $k$'s
are: $k_{I}\!=\!\left\{ 1,0,1,1,1,2,1\right\} $, and $k_{E}\!=\!\left\{
6,5,4,4\right\} $. Thus, this configuration contributes $1,5,1$ to $\protect%
\rho _{I}$ ($k\!=\!0,1,2$) and $2,1,1$ to $\protect\rho _{E}$ ($k\!=\!4,5,6$),
respectively.}
\label{nodes}
\end{figure}

The underlying dynamics of these systems violates detailed balance. As a 
result, the stationary state will be a non-equilibrium steady state (NESS), 
in the sense that persistent probability currents will prevail \cite{ZS2007}. 
Thus, we encounter serious challenges in our attempts to understand analytically
these dramatically different behaviors. Nevertheless, we believe some
insight into our interacting networks can be gained by turning to limiting
cases which embody the main features of the full system. In this spirit, we
consider the ultimate limit: $\kappa _{1}=0,\kappa _{2}=\infty $. In other
words, these are e\underline{x}treme \underline{i}ntroverts and \underline{e}%
xtroverts (or $XIE$, for short). The rules cannot be simpler: When chosen to
act, an introvert will cut a random existing link, while an extrovert will
add a link to a random individual in the population not already connected. 
To implement Monte Carlo simulations, one significant simplification is 
evident. The two intra-communities will quickly end in a null and a complete 
graph, and only the $I$-$E$ cross-links are dynamic. With all intra-community
links frozen (as illustrated in Fig.~\ref{nodes}), our active network
reduces to bipartite graphs and the configuration space of our system
reduces to incidence matrices, $\mathbb{N}$, i.e., a $N_{I}\times N_{E}$
rectangle in the full adjacency matrix, $\mathbb{A}$. Therefore, instead of
having to consider $2^{N(N-1)/2}$ configurations, we can focus on only 
$2^{\mathcal{N}}$ configurations, where%
\begin{equation}
\mathcal{N}\equiv N_{I}N_{E}
\end{equation}%
is the maximum number of cross-links allowed. Given the evolution rules,
there are only two control parameters, $N_{I}$ and $N_{E}$, in our model.
Though this system may appear to be minimal and devoid of interest, we find
surprising features associated with $X$ as well as the degree distributions.
In particular, there is an extraordinary phase boundary along the 
$N_{I}=N_{E}$ line.

Let us denote an element of $\mathbb{N}$\ by $n_{ij}$, which is $1\,$or $0$
when the link between an introvert node $i$ and an extrovert $j$ is present or
absent, respectively. Clearly, $i\in \left[ 1,N_{I}\right] $ and $j\in \left[
1,N_{E}\right] $. From the microscopic $n_{ij}$, various quantities of
interest can be formed, e.g., $X=\Sigma _{ij}n_{ij}$. At another level, the
degree of an $I$ ($E$) node, can be obtained by summing $n$ along a row
(column). To expose a not-so-explicit symmetry, we consider the \textit{%
complement }of a degree, namely, the number of links (associated with a
particular node) \textit{less} than the maximum possible value (i.e., $N-1-k$).
This measure is especially suited for characterizing an $E$ node. Letting $%
\bar{n}_{ij}\equiv 1-n_{ij}$, we define%
\begin{equation}
k_{i}\equiv \Sigma _{j}n_{ij};~~p_{j}\equiv \Sigma _{i}\bar{n}_{ij}
\label{kpbar}
\end{equation}
which are the degree of an $I$ node $i$ and the complement of the degree of
an $E$ node $j$, respectively. Note that, for the $XIE$ model, $k_{i}\in %
\left[ 0,N_{E}\right] $ and $p_{j}\in \left[ 0,N_{I}\right] $. Meanwhile,
the dynamics correspond to changing an element of $\mathbb{N}$ from $1$ to 
$0$ or vice versa. Mapping this binary variable to $\pm 1$, our model
becomes a kinetic Ising model with spin flip dynamics \cite{Glauber63}. In
the lattice gas language \cite{YangLee52}, $k_{i}$ is the number of particles
in row $i$, while $p_{j}$ is the number of holes in column $j$. Similar
to the Ising case (but with an additional exchange operation), the key
symmetry here is 
\begin{equation}
n_{ij}\Leftrightarrow \bar{n}_{ji}\oplus N_{I}\Leftrightarrow N_{E}
\label{ph-sym}
\end{equation}%
which we will refer to as `particle-hole symmetry.' A layman's way to phrase
this symmetry is: The presence of a link is as intolerable to an introvert
as the absence of one (i.e., presence of a `hole') to an extrovert. This
symmetry will play an important role in discussions below. In particular, it
behooves us to characterize the state of an extrovert by a `hole
distribution'%
\begin{equation}
\zeta \left( p\right) =\rho \left( N-1-p\right)
\end{equation}%
rather than the standard degree distribution, $\rho \left( k\right) $.

\subsection{Master equation and the exact microscopic steady state
distribution}

A complete analytical description of the $XIE$ model is given by $\mathcal{P}%
(\mathbb{N},t~|\mathbb{N}_{0},0)$, which is the probability of finding the
system in configuration $\mathbb{N}$ after $t$ steps, starting with initial
configuration $\mathbb{N}_{0}$. (In simulations, we typically let 
$\mathbb{N}_{0}$ be the null matrix or a randomly half filled one. 
But, since our main interest will be properties of
the stationary state, we will drop the $\mathbb{N}_{0},0$ part.) The
discrete master equation for $\mathcal{P}$ can be written:%
\begin{equation}
\mathcal{P}(\mathbb{N},t+1)-\mathcal{P}(\mathbb{N},t)=\sum_{\{\mathbb{N}%
^{\prime }\}}[W(\mathbb{N},\mathbb{N}^{\prime })\mathcal{P}(\mathbb{N}%
^{\prime },t)-W(\mathbb{N}^{\prime },\mathbb{N})\mathcal{P}(\mathbb{N},t)]
\label{ME}
\end{equation}%
where $W(\mathbb{N},\mathbb{N}^{\prime })$ is the probability for
configuration $\mathbb{N}^{\prime }$ to become $\mathbb{N}$ in a step
(attempt). Using (\ref{kpbar}), $W\left( \mathbb{N}^{\prime },\mathbb{N}%
\right) $ can be easily written: 
\begin{equation}
\sum\limits_{i,j}\frac{\Delta }{N}\left[ \frac{\Theta \left( k_{i}\right) }{%
k_{i}}\bar{n}_{ij}^{\prime }n_{ij}+\frac{\Theta \left( p_{j}\right) }{p_{j}}%
n_{ij}^{\prime }\bar{n}_{ij}\right]  \label{rates}
\end{equation}%
where $\Theta \left( x\right) $ is the Heavyside function (i.e., $1$ if $x>0$
and $0$ if $x\leq 0$) and $\Delta \equiv \Pi _{k\ell \neq ij}\delta \left(
n_{k\ell }^{\prime },n_{k\ell }\right) $ ensures that only $n_{ij}$ may
change in a step. With such a random sequential scheme, each node has an
even chance of being chosen after $N$ attempts. Thus, $N$ attempts is often
referred to as one Monte Carlo step (MCS), so that a run of $\tau $ MCS
involves $\tau N$ `spin-flip attempts.'

The dynamics defined here is clearly ergodic. More remarkable is that it
obeys detailed balance, as shown in Appendix A. Consequently, in the 
$t\rightarrow \infty $ limit, $\mathcal{P}$ approaches a unique stationary
distribution, $\mathcal{P}^{ss}$, while all probability currents vanish. More
crucially, detailed balance allows us to find $\mathcal{P}^{ss}$ by applying 
$\mathcal{P}^{ss}\left( \mathbb{N}\right) =\mathcal{P}^{ss}\left( \mathbb{N}%
^{\prime }\right) W\left( \mathbb{N},\mathbb{N}^{\prime }\right) /W\left( 
\mathbb{N}^{\prime },\mathbb{N}\right) $ repeatedly. Imposing normalization,
we arrive at an explicit closed form: 
\begin{equation}
\mathcal{P}^{ss}\left( \mathbb{N}\right) =\frac{1}{\Omega }%
\prod\limits_{i=1}^{N_{I}}\left( k_{i}!\right)
\prod\limits_{j=1}^{N_{E}}\left( p_{j}!\right)  \label{P*}
\end{equation}%
where $\Omega =\Sigma _{\left\{ \mathbb{N}\right\} }\Pi \left( k_{i}!\right)
\Pi \left( p_{j}!\right) $ is a `partition function.' Note that the
particle-hole symmetry (c.f. Eqn.~\ref{ph-sym}) is manifest here.

Interpreting $\mathcal{P}^{ss}$ as a Boltzmann factor, we can write a
`Hamiltonian'
\footnote{Of course, $\ln \left( \Sigma _{j}n_{ij}\right) !$ can be cast as 
$\Sigma_{\ell }\ln \left( \Sigma _{j}n_{ij}-\ell \right) $ but this form is 
hardly a simplification.} 
\begin{equation}
\mathcal{H}\left( \mathbb{N}\right) =-\left\{ \sum\limits_{i=1}^{N_{I}}\ln
\left( \sum\limits_{j=1}^{N_{E}}n_{ij}\right) !+\sum\limits_{j=1}^{N_{E}}\ln
\left( \sum\limits_{i=1}^{N_{I}}\bar{n}_{ij}\right) !\right\}  \label{H}
\end{equation}%
Now, this form immediately alerts us to the level of complexity of this
system of `Ising spins,' as $\mathcal{H}$ contains a peculiar form of long
range interactions. Each `spin' is coupled to all other `spins' \textit{in
its row and column}, via all possible types of `multi-spin' interactions! 
We are not aware of any system in solid state physics with this kind of
interactions. Yet, $\mathcal{H}$ exposes the underlying structure of (a
limit of) a very simple model of social interactions. Meanwhile, it is
understandable that computing $\Omega $, let alone statistical properties of
macroscopic quantities, will be quite challenging. Nevertheless, as the next
two sections show, we are able to exploit mean-field approaches to predict
various quantities, for generic points in the space of control parameters: $%
\left( N_{I},N_{E}\right) $. As in standard equilibrium statistical systems,
our mean-field theory fails in the neighborhood of critical points, which
turn out to be the $N_{I}=N_{E}$ line here.

\section{Degree distributions: simulation results and theoretical
considerations}

In this section, we focus on a quantity most commonly studied in networks:
the degree distribution, $\rho (k)$. Of course, there are several types of 
$\rho $'s for a system with $G$ subgroups or communities. If $k_{AB}$ denotes
the number of nodes in group $B$ connected to a node in group $A$ (i.e., the
links $A$ has to $B$), then, in general, $\left\langle k_{AB}\right\rangle
\neq \left\langle k_{BA}\right\rangle $. Instead, we have $N_{A}\left\langle
k_{AB}\right\rangle =N_{B}\left\langle k_{BA}\right\rangle $ as the average
number of all the links between subgroups $A$ and $B$, while the numbers of
nodes in each, $N_{A}$ and $N_{B}$, are not necessarily equal. Thus, we have 
$G^{2}$ degree distributions in general. For $XIE$ however, the internal
links, $k_{AA}$, are static with trivial distributions (e.g., $\delta \left(
k_{II}\right) $). Thus, it is sufficient to describe each subgroup by just
one degree distribution, associated with $k_{IE}$ and $k_{EI}$. For
simplicity, we will denote these as $\rho _{I}(k_{I})$ and $\rho _{E}(k_{E})$,
respectively (e.g., $\rho _{I}(1)=5$ and $\rho _{E}(6)=1$ in Fig.~\ref{nodes}). 
In view of the particle-hole symmetry discussed above, we will
often consider $\zeta _{E}\left( p_{E}\right) $ instead.

\subsection{Simulation results for $\protect\rho _{I}$ and $\protect\rho %
_{E} $}

Our main goal is to demonstrate that interesting phenomena can emerge from a
minimal model like $XIE$, and that they can be understood reasonably well by
mean-field analysis, rather than a systematic and exhaustive study of this
particular system. Thus, we restrict our simulation studies mainly to 
$N=N_{I}+N_{E}=200$, a number large enough to show collective behavior, yet
small enough for gathering good statistics. Starting with a network with 
various initial conditions (null graph, complete graph, random half-filled), 
we evolve the system according to the simple rules given above. Not
surprisingly, after $O\left( N\right) $ MCS, all the $I$-$I$ links are absent
while all the $E$-$E$ links are present. To be quite certain that the system has
equilibrated, we discard the first $5\times 10^{7}$ MCS. Thereafter, we 
measure the degrees of each node every $50$ MCS. The distributions are then 
obtained as the average over $10^{8}$ measurements. As our main focus is not 
the transient behavior, but rather only steady state properties, we will drop
the superscripts $ss$ to simplify notation.

\begin{figure*} 
  \centering
  \mbox{
    \subfigure{\includegraphics[width=3.25in]{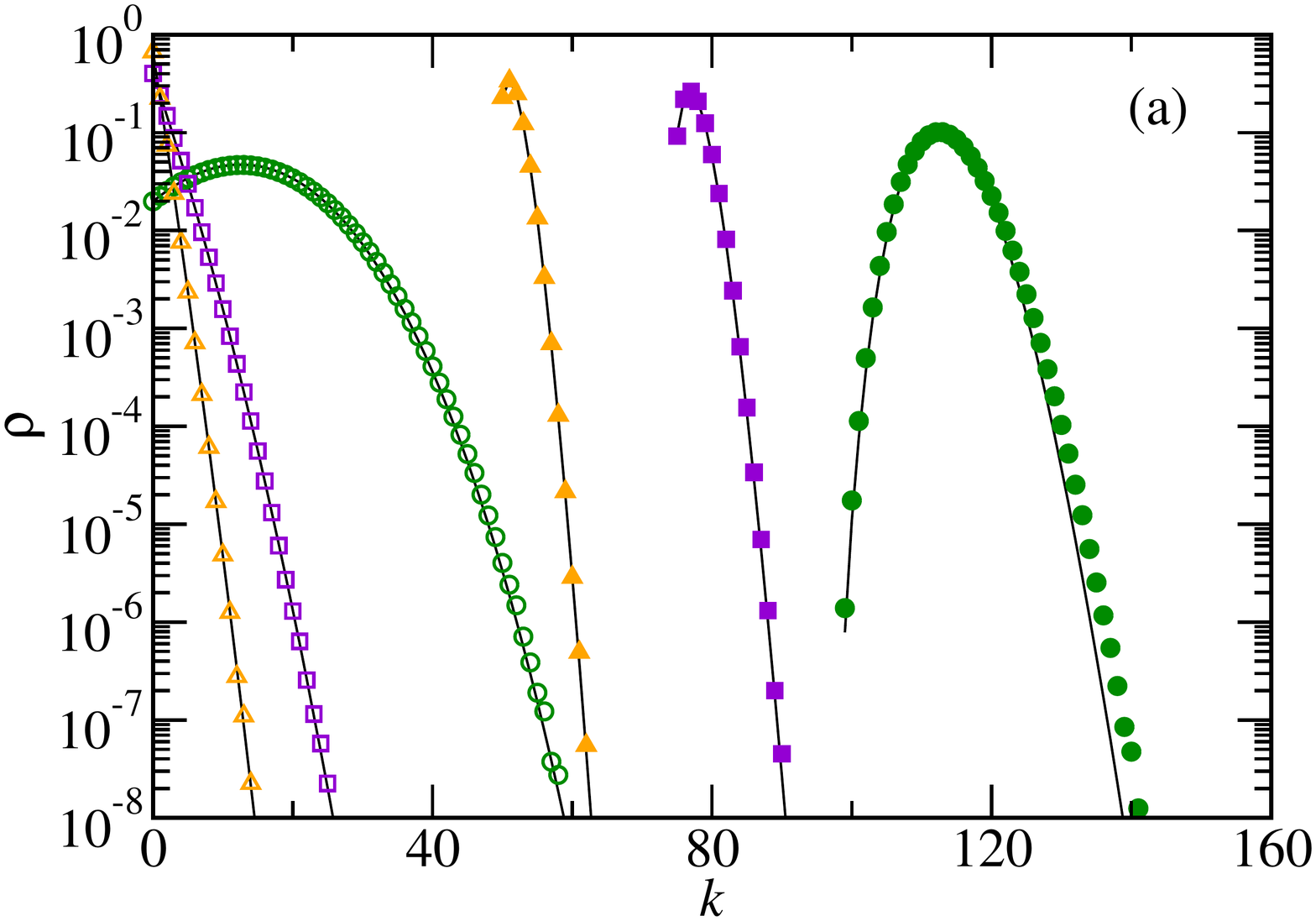}}\quad
    \subfigure{\includegraphics[width=3.25in]{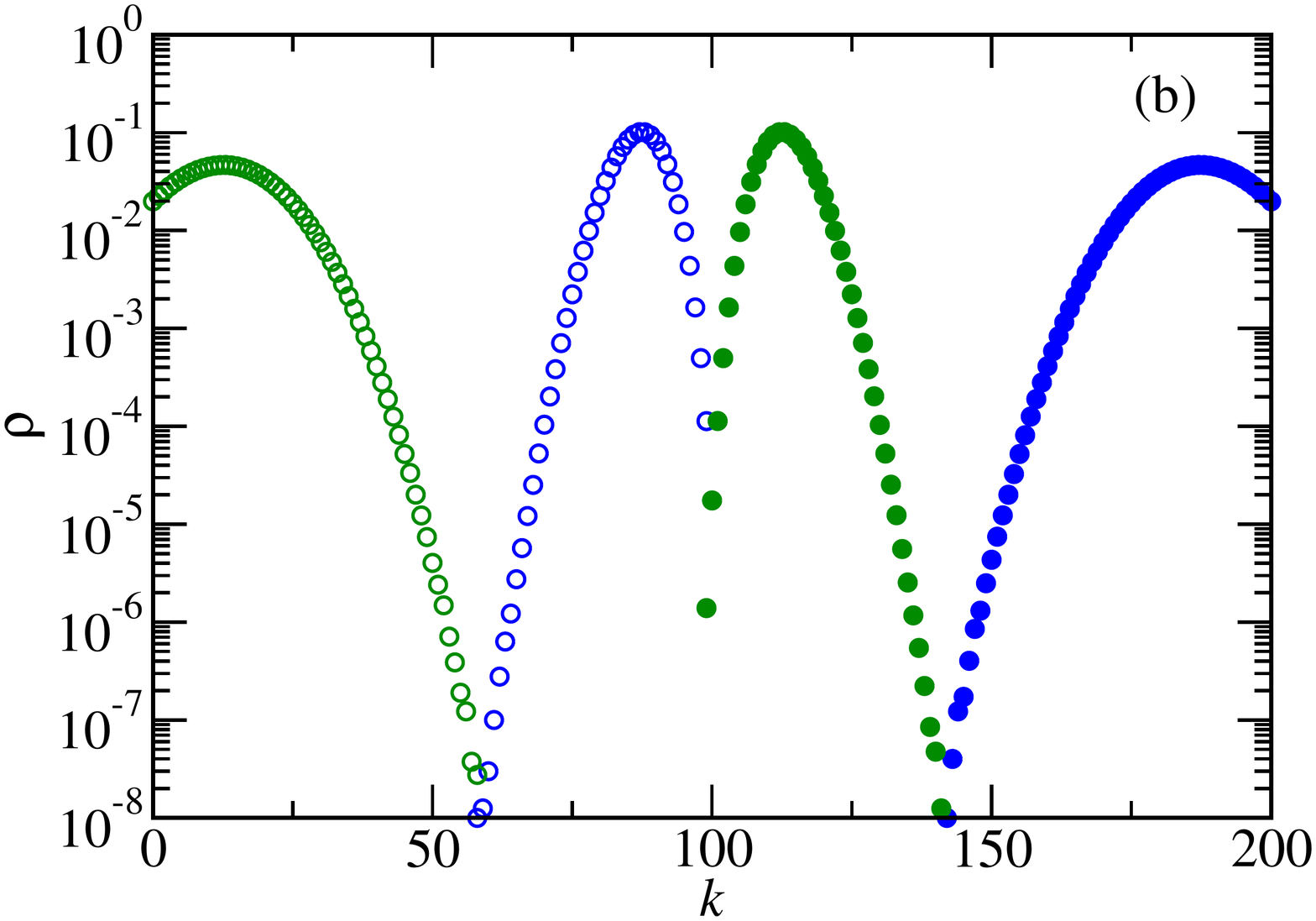}}
     }\vspace{-0.8cm}
  \caption{Degree distributions, $\rho $, for several cases with $%
N_{I}+N_{E}=200$. Simulation results for the low/high $k$ components,
associated with introverts/extroverts, are denoted by open/solid symbols.
(a) The symbols for $(N_{I},N_{E})$ are orange triangles $(150,50)$, purple
squares $(125,75)$, and green circles $(101,99)$. The solid black lines are
predictions from a self-consistent mean-field theory. (b) When two
introverts `change sides,' a dramatic jump in $\rho \left( k\right) $
results, with the case of $(99,101)$ shown as blue circles.}
\label{DD}
\end{figure*}

The degree distributions, $\rho \left( k\right) $, for three cases -- $%
\left( N_{I},N_{E}\right) =\left( 150,50\right)$, $\left( 125,75\right)$, $\left( 101,99\right) $ -- are shown in Fig.~\ref{DD}(a). Evidently, each 
$\rho $ consists of two disjoint components, associated with small/large $k$.
It is clear that, since the $I$'s are only linked to the $E$'s, the
distribution associated with $k\leq N_{E}$ can be attributed to $\rho
_{I}(k_{I})$. Similarly, since all $E$-$E$ links are present, $\rho
_{E}(k_{E})$ can be identified with the component with $k\geq N_{E}-1$. For
these cases, there is no overlap at $k=N_{E}-1,N_{E}$ so that we can label 
$\rho _{I}$ and $\rho _{E}$ without ambiguity: open and solid symbols
respectively. Apart from having these two components, the most prominent
feature is that neither component resembles $3^{-\left\vert k-\kappa
\right\vert }$, the degree distribution of a homogeneous population with
preferred degree $\kappa $ \cite{CSP24,LiuJoladSchZia13}. As will be shown,
they are well approximated by Poisson distributions, the analytic result of
a mean-field approach.

Of course, the averages obey $\langle k_{I}\rangle <\langle k_{E}\rangle $,
especially since they must satisfy%
\begin{equation}
\langle k_{I}\rangle N_{I}=(\langle k_{E}\rangle -N_{E}+1)N_{E}  \label{kIkE}
\end{equation}%
which is the average of the total number of cross-links: $\langle X\rangle $.
Meanwhile, we can expect that, since each node is given the same chance to
act, all these averages will increase as the ratio $N_{E}/N_{I}$ increases.
Other than these obvious aspects, a casual glance shows several surprising
features. For example, $\langle k_{I}\rangle \cong 0.5$, $1.4$, and $13.9$
for the $\rho _{I}$'s in Fig.~\ref{DD}(a), associated with, respectively, $%
N_{E}/N_{I}=1/3$, $3/5$, and $99/101$. Naively, $N_{E}/N$ and $N_{I}/N$ may
be thought of as the probability of adding and cutting a cross-link, leading
to the expectation that the ratio $\left\langle X\right\rangle /\left[ 
\mathcal{N-}\left\langle X\right\rangle \right] $ (presence/absence of link)
is $N_{E}/N_{I}$. A little algebra leads further to the prediction $\langle
k_{I}\rangle =N_{E}^{2}/N$, i.e., approximately $12.5$, $28.1$, and $49.0$.
Needless to say, this naive picture is far from realized.

Before considering a more successful theoretical approach, we present the
system's behavior for $N_{I}<N_{E}$. In Fig.~\ref{DD}(b), we illustrate the
degree distribution for $(N_{I},N_{E})=(99,101)$ (blue circles), as well as
the previous case of $(101,99)$ (green circles). In other words, just two
introverts have `changed sides' here. The first remarkable feature is the
sizeable jump, easily discerned by focusing on say, the two $\rho _{I}$'s,
shown with open symbols.
The other notable feature is the symmetry, which can be traced to the
underlying `particle-hole symmetry.' By exchanging $N_{I}\Leftrightarrow
N_{E}$ and plotting the degree distribution \textit{vs}. $p\equiv N-1-k$, we
find excellent overlap between the blue and green data points. Of course,
such a plot displays $\zeta \left( p\right) $, the `hole distribution.'

\begin{figure}[tbp]
\centering
\includegraphics[width=3.5in]{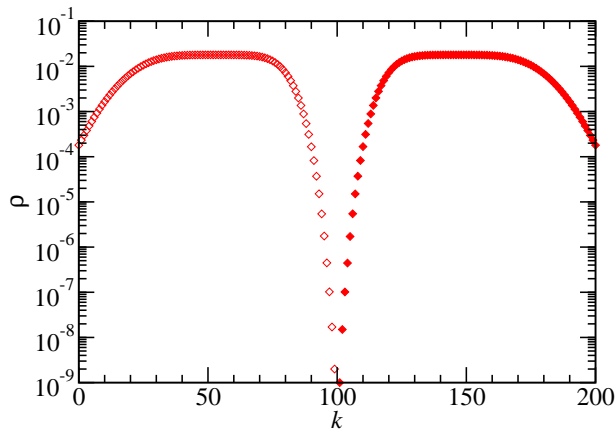}\vspace{-0.8cm}
\caption{Simulation results of the degree distribution for the symmetric $%
\left( 100,100\right) $ case. The two components, denoted by open and solid
diamonds, can be associated with the separate distributions 
$\protect\rho _{I}$ and $\protect\rho _{E}$, respectively.}
\label{100DD}
\end{figure}

Finally, let us turn to the symmetric case $(100,100)$, which appears to be
most challenging, for both Monte Carlo simulations and theoretical
understanding. First of all, the run time it takes for the system to settle
is much longer, typically a hundred-fold longer than the $N_{I}\neq N_{E}$
cases. To compile a reliable histogram for the $\rho \left( k\right) $,
shown in Fig.~\ref{100DD}, we take $10^{10}$ measurements in a combination
of $5$ runs, each of which lasts for $10^{11}$ MCS (after discarding the 
first $10^7$ MCS for the system to settle into steady states). Note that, 
to untangle $\rho _{I,E}$ in the
central pair of points, we recorded separately whether an $I$ or a $E$ node
has $99,100$ links. Of course, the distribution is symmetric, i.e., 
$\rho_{I}=\zeta _{E}$. The unexpected aspect is the presence of broad and flat
plateaux, quantitative understanding of which remains elusive. As will be
discussed in Section $4$, models with $N_{I}=N_{E}$ can be regarded as
critical systems in which large fluctuations and critical slowing down can
be typically expected. In the next subsection, we will sidestep this
problematic case and focus only on systems with $N_{I}\neq N_{E}$, for which
a mean-field approach proves to be quite successful.

\subsection{Self-consistent mean-field approximation}

Given the exact steady state distribution (Eqn.~(\ref{P*})), the $\rho $'s can
be computed, in principle, via%
\begin{eqnarray}
\rho _{I}\left( k_{I}\right) &=&\sum_{\left\{ \mathbb{N}\right\} }\delta
\left( k_{I}-\Sigma _{j}n_{ij}\right) \mathcal{P}^{ss}\left( \mathbb{N}%
\right) \\
\rho _{E}\left( k_{E}\right) &=&\sum_{\left\{ \mathbb{N}\right\} }\delta
\left( k_{E}-\Sigma _{i}n_{ij}\right) \mathcal{P}^{ss}\left( \mathbb{N}%
\right)
\end{eqnarray}%
(for any $i$ in the first equation and any $j$ in the second). 
In practice, this task is extremely challenging. In particular, since $k_{I,E}$
correspond to, in a 2-dimensional Ising model, the total mangetization in a
row or a column, it is understandable that computing their full 
\textit{distributions} is beyond our reach. Thus, we resort to a mean-field
approach, applied to the underlying \textit{dynamics} of the model 
\cite{LiuJoladSchZia13,LiuSchZia14}. In other words, we implement an
approximation scheme on the transition probabilities for the degree of a 
\textit{particular} node to increase/decrease by unity: $R\left(
k\rightarrow k\pm 1\right) $. Once these are determined, the \textit{approximate}
(signified with a tilde above) steady state degree distribution must satisfy%
\begin{equation}
\tilde{\rho}\left( k\right) R\left( k\rightarrow k-1\right) =\tilde{\rho}%
\left( k-1\right) R\left( k-1\rightarrow k\right)  \label{RR-rho}
\end{equation}%
and can be found in closed form.

Specifically, we first consider a particular $I$ node, with $\rho _{I}\left(
k_{I}\right) $ being the probability to find it having $k_{I}$ links. Then,
provided $k_{I}>0$, $R_{I}\left( k_{I}\rightarrow k_{I}-1\right) =1/N$ which
is the probability that this node is chosen to act. By contrast, the exact
rate for having a link added ($k_{I}-1\rightarrow k_{I}$) is more
complicated, since it depends not only on all the $N_{E}-k_{I}+1$ extroverts 
\textit{not} connected to it, but also on how many `holes' each has --
through $1/p_{j}$ (in Eqn.~(\ref{rates})). To proceed, we must make judicious
approximations. In the spirit of mean-field theory, we can replace $1/p_{j}$
by the average $\left\langle 1/p_{E}\right\rangle ^{\prime }$, where the
prime stands for an average \textit{restricted} to nodes with $p_{E}>0$.
Though we can formulate the theory with $\left\langle 1/p_{E}\right\rangle
^{\prime }$, let us make a further simplifying approximation and replace it
by $1/\left\langle p_{E}\right\rangle ^{\prime }$. So, we write%
\begin{equation}
R_{I}\left( k_{I}-1\rightarrow k_{I}\right) \cong \frac{N_{E}-k_{I}+1}{N}%
\frac{1}{\left\langle p_{E}\right\rangle ^{\prime }}
\end{equation}%
If we had the distribution of an extrovert's holes, $\zeta _{E}\left(
p_{E}\right) $, then we have the following relation:%
\begin{equation}
\left\langle p_{E}\right\rangle ^{\prime }\equiv \frac{\sum_{p_{E}>0}p_{E}%
\zeta _{E}\left( p_{E}\right) }{\sum_{p_{E}>0}\zeta _{E}\left( p_{E}\right) }%
=\frac{\langle p_{E}\rangle }{1-\zeta _{E}\left( 0\right) }  \label{p'}
\end{equation}%
Of course, since $\rho _{E}\left( k_{E}\right) $ is unknown, so is $\zeta
_{E}\left( p_{E}\right) $. As will be shown, the goal of a self-consistent
mean-field theory is to find an approximate expression for these
distributions, as well as for $\rho _{I}$.

Proceeding, we exploit Eqn.~(\ref{RR-rho}) and readily find 
\begin{eqnarray}
\tilde{\rho}_{I}\left( k_{I}\right) &=&\frac{N_{E}-k_{I}+1}{\left\langle
p_{E}\right\rangle ^{\prime }}\frac{N_{E}-k_{I}+2}{\left\langle
p_{E}\right\rangle ^{\prime }}...\frac{N_{E}}{\left\langle
p_{E}\right\rangle ^{\prime }}\tilde{\rho}_{I}\left( 0\right)  \notag \\
&\propto &\frac{\left( \left\langle p_{E}\right\rangle ^{\prime }\right)
^{N_{E}-k_{I}}}{\left( N_{E}-k_{I}\right) !}  \label{rhoI}
\end{eqnarray}%
Since $p_{I}\equiv N_{E}-k_{I}$ is the number of `holes' associated with an 
$I$ node, we recognize this as a Poisson distribution (truncated at $N_{E}$)
for the hole distribution. Imposing normalization, we find a compact closed
from, $\tilde{\zeta}_{I}\left( p_{I}\right) =\left( \left\langle
p_{E}\right\rangle ^{\prime }\right) ^{p_{I}}/Z_{I}p_{I}!$, where $%
Z_{I}=\Sigma _{\ell =0}^{N_{E}}\left( \left\langle p\right\rangle ^{\prime
}\right) ^{\ell }/\ell !$ is the sum of the first $N_{E}+1$ terms of an
exponential series. Despite its simplicity, we feel that the notation 
$\tilde{\zeta}_{I}\left( p_{I}\right) $ may be too confusing and so, 
we will remain with $\tilde{\rho}_{I}$ instead:%
\begin{equation}
\tilde{\rho}_{I}\left( k_{I}\right) =\frac{\left( \left\langle
p_{E}\right\rangle ^{\prime }\right) ^{N_{E}-k_{I}}}{Z_{I}\left(
N_{E}-k_{I}\right) !}  \label{rho-tilde}
\end{equation}

Of course, $\left\langle p_{E}\right\rangle ^{\prime }$ is still an unknown
parameter at this point. For that, we turn to a particular $E$ node and,
exploiting `particle-hole' symmetry, consider its hole distribution, $\zeta
_{E}\left( p_{E}\right) $. Since adding a link is decreasing $p_{E}$ by
unity, we again have $R_{E}\left( p_{E}+1\rightarrow p_{E}\right) =1/N$, the
probability that this node is chosen to act, provided $p_{E}>0$. Meanwhile,
it is connected to $N_{I}-p_{E}$ (i.e., $k_{E}-N_{E}+1$) introverts, each of
which has $k_{i}$ links. As above, we rely on the same arguments and
replace the $k_{i}$'s by a suitable average:%
\begin{equation}
R_{E}\left( p_{E}\rightarrow p_{E}+1\right) \cong \frac{N_{I}-p_{E}}{N}\frac{%
1}{\left\langle k_{I}\right\rangle ^{\prime }}
\end{equation}%
where 
\begin{equation}
\left\langle k_{I}\right\rangle ^{\prime }=\frac{\langle k_{I}\rangle }{%
1-\rho _{I}\left( 0\right) }  \label{k'}
\end{equation}%
Recasting Eqn.~(\ref{RR-rho}) for $\tilde{\zeta}$, we have
\begin{equation}
\tilde{\zeta}_{E}\left( p_{E}\right) =\frac{\left\langle k_{I}\right\rangle
^{\prime }}{N_{I}-p_{E}}\tilde{\zeta}_{E}\left( p_{E}+1\right)
\end{equation}%
This recursion relation leads to a (truncated) Poisson distribution in 
$N_{I}-p_{E}$, and imposing normalization, we have 
\begin{equation}
\tilde{\zeta}_{E}\left( p_{E}\right) =\frac{\left( \left\langle
k_{I}\right\rangle ^{\prime }\right) ^{N_{I}-p_{E}}}{Z_{E}\left(
N_{I}-p_{E}\right) !}  \label{zeta-tilde}
\end{equation}%
with $Z_{E}=\Sigma _{\ell =0}^{N_{I}}\left( \left\langle k\right\rangle
^{\prime }\right) ^{\ell }/\ell !$. Of course, we recognize $N_{I}-p$ is
just the number of \textit{cross}-links associated with a $E$ node: 
$k_{E}-N_{E}+1$. Thus, the expression for 
$\tilde{\rho}_{E}\left(k_{E}\right) $ will not be simpler. Note that, 
along with Eqn.~(\ref{rho-tilde}), this result again confirms the underlying 
particle-hole symmetry.

Though $\left\langle k_{I}\right\rangle ^{\prime }$ is also an unknown, we
can compute both it and $\left\langle p_{E}\right\rangle ^{\prime }$ using
the approximate distributions $\tilde{\rho}_{I}$ and $\tilde{\zeta}_{E}$ in
Eqns.~(\ref{k'},\ref{p'}) instead. Since $\tilde{\rho}_{I}$ and $\tilde{\zeta%
}_{E}$ depend on $\left\langle p_{E}\right\rangle ^{\prime }$and $%
\left\langle k_{I}\right\rangle ^{\prime }$, respectively, we may define the
functions $f$ and $g$:%
\begin{eqnarray}
\left\langle k_{I}\right\rangle ^{\prime }\cong \frac{\Sigma k_{I}\tilde{\rho%
}_{I}\left( k_{I}\right) }{1-\tilde{\rho}_{I}\left( 0\right) }\equiv f\left(
\left\langle p_{E}\right\rangle ^{\prime }\right);\nonumber \\
~~\left\langle
p_{E}\right\rangle ^{\prime }\cong \frac{\Sigma p_{E}\tilde{\zeta}_{E}\left(
p_{E}\right) }{1-\tilde{\zeta}_{E}\left( 0\right) }\equiv g\left(
\left\langle k_{I}\right\rangle ^{\prime }\right)
\end{eqnarray}%
Making a plot of these functions in the $\left\langle k\right\rangle
^{\prime }$-$\left\langle p_{E}\right\rangle ^{\prime }$ plane, the point of
intersection then determines, self-consistently, the values for these two
parameters. In practice, it is simple to start with, say, a trial value $%
p_{0}$ for $\left\langle p_{E}\right\rangle ^{\prime }$ and compute $%
\left\langle k_{I}\right\rangle ^{\prime }$ through Eqn.~(\ref{rho-tilde}).
Inserting this $\left\langle k_{I}\right\rangle ^{\prime }$ into Eqn.~(\ref
{zeta-tilde}), we compute $\tilde{\zeta}_{E}$ and the associated $%
\left\langle p_{E}\right\rangle ^{\prime }$. If this result is not $p_{0}$,
then vary the latter until they agree. In other words, this process will
find the solution to $\left\langle p_{E}\right\rangle ^{\prime }=g\left(
f\left( \left\langle p_{E}\right\rangle ^{\prime }\right) \right) $. Instead
of quoting the self consistent values for $\left\langle p_{E}\right\rangle
^{\prime }$and $\left\langle k_{I}\right\rangle ^{\prime }$, we plot the
full distributions predicted by Eqns.~(\ref{zeta-tilde},\ref{rho-tilde}), shown
as solid black lines in Fig.~\ref{DD}(a). We should emphasize that \textit{no%
} fit parameters have been introduced in this approach; the lines depend
only on the control parameters, $N_{I,E}$. It is clear that the agreement
between theory and simulation data is excellent for $N_{I}/N_{E}\gg 1$. By
symmetry, it will also be quite good for cases with $N_{I}\ll N_{E}$. For
the $\left( 101,99\right) $ case, disagreement between theory and data is
visibly detectable, a sign that correlations are no longer negligible. We
did not plot the predictions for the $N_{I}=N_{E}=100$ case, as the theory
is, not surprisingly, deficient.

While there is reasonably good agreement between this theory and simulation
data for systems with $N_{I}\neq N_{E}$, it does not offer much insight
into the dramatic changes when just two individuals `change sides,' nor the
emergence of broad plateau in Fig.~\ref{100DD}. To address some of these
issues, we turn next to a more macroscopic perspective,
focusing only on the total number of cross-links, $X$.

\begin{figure}[tbp]
\centering
\includegraphics[width=3.5in]{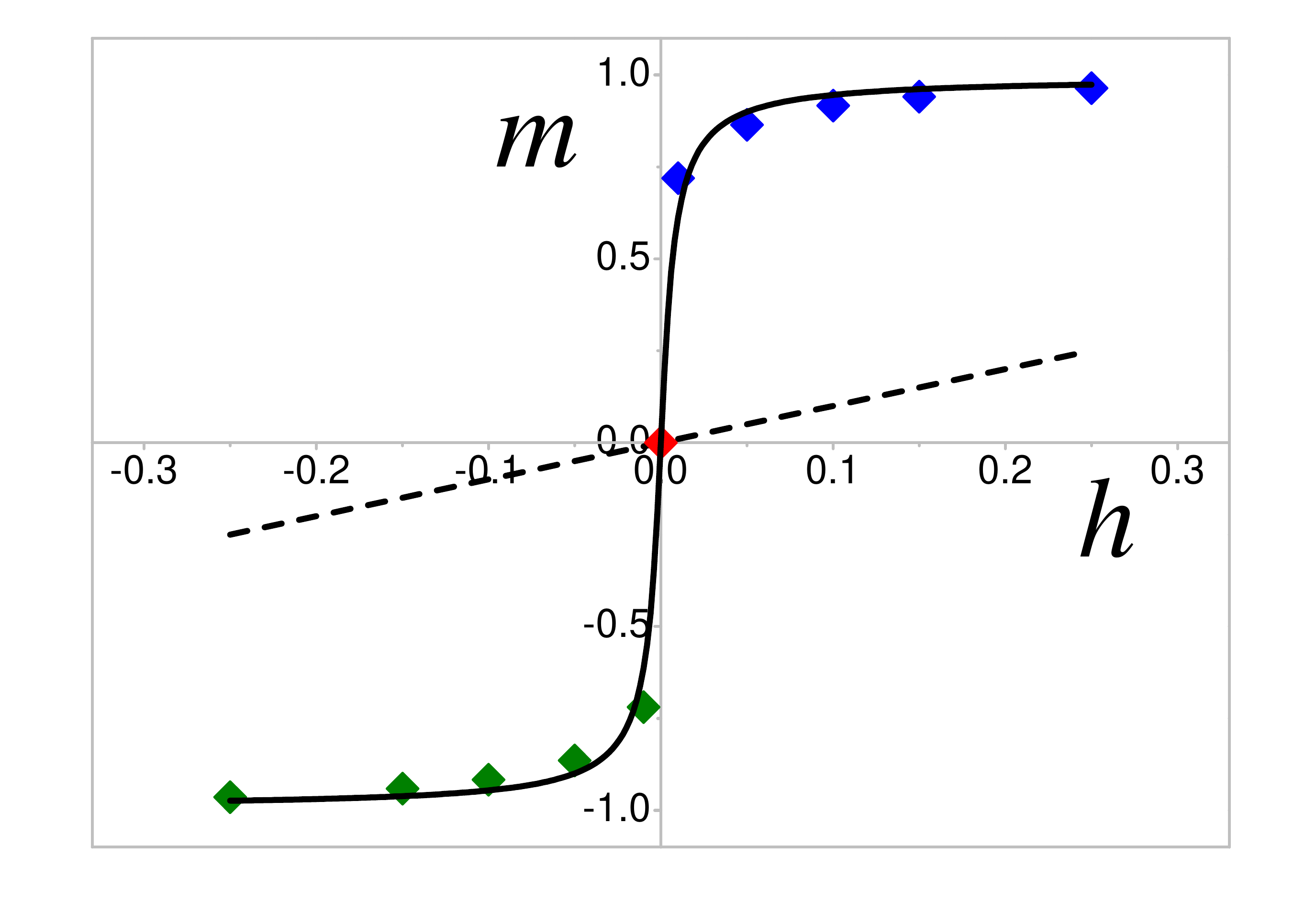}\vspace{-0.8cm}
\caption{ The behavior of $\left\langle X\right\rangle $ for various systems
with $N_{I}+N_{E}=200$, displayed in terms of $m\left( h\right) $. Green
(blue) diamonds are associated with $N_{I}$ ($N_{E}$) being $%
125,115,110,105, $ and $101.$ The red diamond is the symmetric, critical
case of $(100,100)$. The dashed line is the prediction from an `intuitively
reasonable'\ argument. A mean-field approach leads to the solid (black)
line. }
\label{mh}
\end{figure}

\section{Statistical properties of $X$, the total number of cross-links}

\begin{figure}[tbp]
\centering
\includegraphics[width=3.5in]{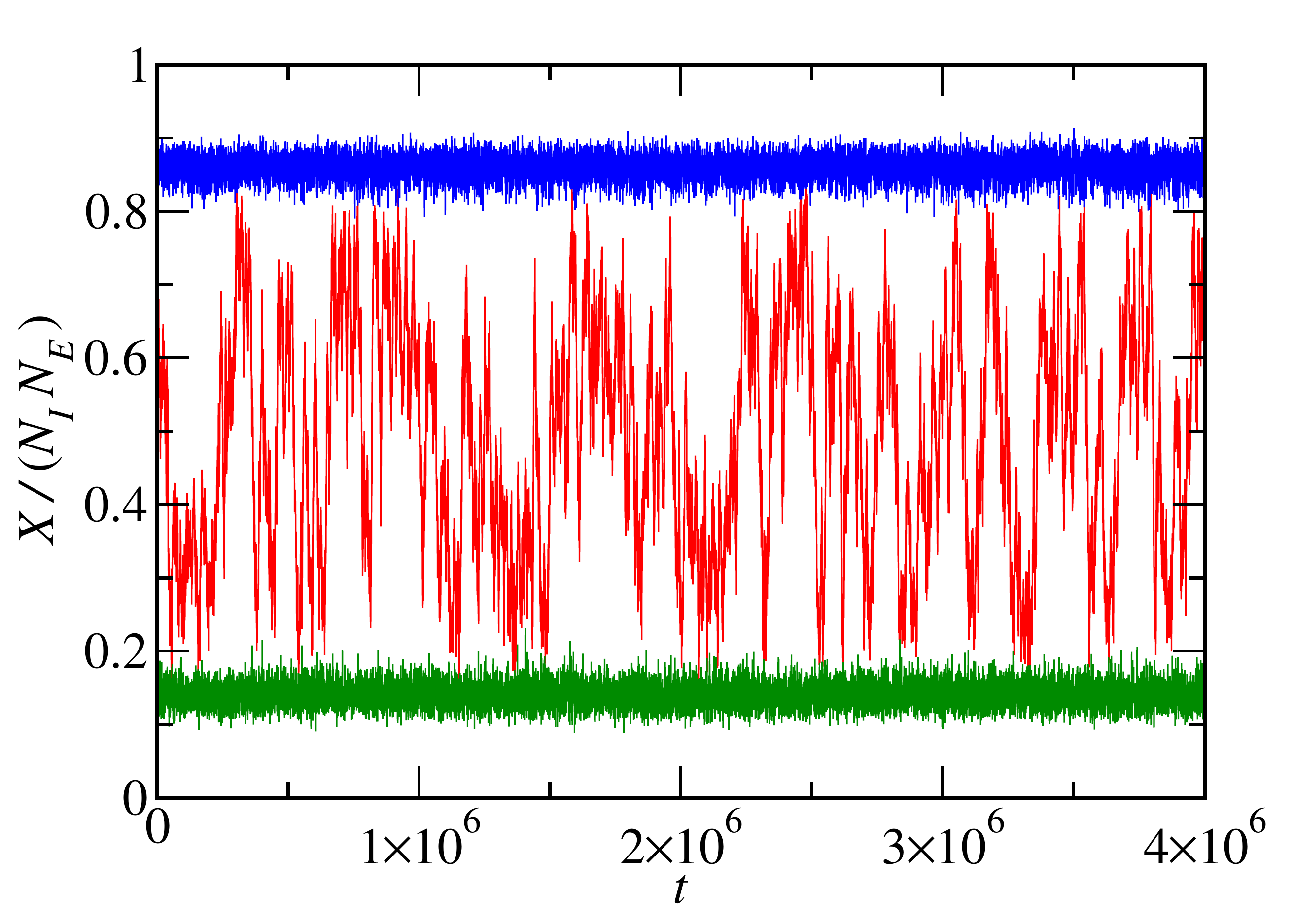}\vspace{-0.5cm}
\caption{Time traces of $X$ for three cases with $\left(
N_{I},N_{E}\right)
$ near/at criticality, shown in green $\left( 101,99\right) $, red $\left(
100,100\right) $, and blue $\left( 99,101\right)$. To clarify, these are
traces in the steady state, with $t=0$ here being $\thicksim 10^8$ MCS
after the start of the runs.}
\label{timetrace}
\end{figure}

Though degree distributions are standard for characterizing networks, the
puzzling features displayed in the case $N_{I}=N_{E}=100$ provide the
motivation to study a simpler quantity. In $XIE$, a natural quantity is the
total number of cross-links, $X$. Since it is a single number, studying $X$
should be considerably easier. Furthermore, through the mapping to the Ising
model, we recognize $2X-\mathcal{N}$ is the total magnetization $M$. Since
the properties of $M$, especially its singular behavior near the critical
point, have been extensively investigated, we believe that, by exploring a
similar quantity in our system, we can benefit from known results. Unlike
the Ising model however, there is essentially only one control parameter in $%
XIE$: $N_{E}-N_{I}$. The other, $\mathcal{N}=N_{E}N_{I}$ or $N=N_{E}+N_{I}$,
should be regarded as the system size, suitable for finite-size scaling
analysis. Of course, since we have defined a `Hamiltonian' (Eqn.~(\ref{H})), it
would be natural to introduce a `temperature' and consider $\mathcal{P}$'s
like $\exp \left[ -\mathcal{H}/k_{B}T\right] $ and perhaps some symmetry
breaking variable as well. Such interesting questions are beyond the scope
of this paper and will be pursued elsewhere. Here, let us focus on $X$ and
its response to $N_{E}-N_{I}$. By defining `intensive' variables%
\begin{eqnarray}
m &\equiv &2\left\langle X\right\rangle /\mathcal{N}-1 \\
h &\equiv &(N_{E}-N_{I})/(N_{E}+N_{I})
\end{eqnarray}%
we can pose our question as: What is the `equation of state' $m\left( h\right) $
for the $XIE$ model?

\subsection{An extraordinary phase transition: Results from simulations and
mean-field theory}

To answer the question posed above, we can use the $\left\langle
k_{I}\right\rangle $ from the simulation data above to construct $%
\left\langle X\right\rangle $ for the introverts and extroverts. Plotting the results in
terms of $m\left( h\right) $, shown as solid diamonds in Fig.~\ref{mh}, we
see that it deviates far from the naive expectation discussed above, namely, $%
\langle k_{I}\rangle =N_{E}^{2}/N$, corresponding to $m=h$ and shown as the
dashed black line. More remarkably, our $m\left( h\right) $\ displays a
sizeable jump -- $70\%$ of the full range when just two individuals `change
sides' (i.e., $\left( 101,99\right) \rightarrow \left( 99,101\right) $).
Though such jumps resemble the equation of state for an Ising model with $%
T\ll T_{c}$, many aspects of our system do not display the typical
characteristics of a first order transition, e.g., hysteresis, finite
fluctuations, etc. The rest of this Subsection is devoted to different ways
we probe this extraordinary transition. As we will see, despite the
appearance of a discontinuity in $m\left( h\right) $, it is reasonable to
refer to the $h\thicksim 0$ region as `critical,' as typical behavior here
conforms to that in systems displaying mixed-order transitions. Though
discovered some time ago, such behavior is far less widely known as ordinary
first and second order phase transitions \footnote{%
An excellent summary of the Thouless effect can be found in ref.\cite%
{BarMukamel14}.}. 

\subsubsection{Time traces, histograms, and power spectra}

\begin{figure}[tbp]
\centering
\includegraphics[width=3.5in]{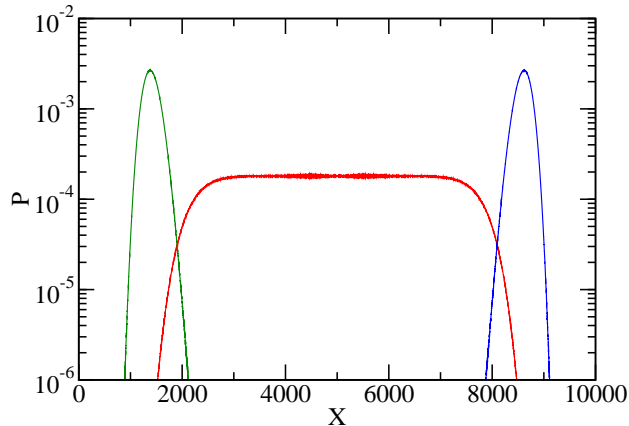}\vspace{-0.8cm}
\caption{The distribution, $P\left( X\right) $, compiled
with the time traces in Fig.~\protect\ref{timetrace}.}
\label{histogram}
\end{figure}

Since we collect time traces in a standard simulation, we should exploit the
information they contain. Illustrated in Fig.~\ref{timetrace} is $X\left(
t\right) $ for the critical system, as well as the two neighboring cases: $%
h=0,\pm 0.01$. Clearly, the trace for $N_{I}=N_{E}=100$ (red line) is
dramatically different from the other two. For $N_{I}=N_{E}\pm 2$, $X$
settles down very quickly, hovering around the average $\langle X\rangle $
with fluctuations of $O(100)$ (i.e., $O(\sqrt{\mathcal{N}})$ here). By
contrast, in the critical case, $X$ wanders widely (i.e., $O(\mathcal{N})$
here) and evolves exceedingly slowly. These time traces can be used to
compile histograms in $X$, from which we obtain the steady state
distribution $P\left( X\right) $, shown in Fig.~\ref{histogram}. Not
surprisingly, they are sharply peaked and Gaussian like for the off-critical
cases (green and blue lines), while the distribution in the $N_{I}=N_{E}$
case (red line) is essentially flat over most of the full range, $\left[ 0,%
\mathcal{N}\right] $. Both the time trace of the critical case and the flat
plateau in $P\left( X\right) $ give the impression of an \textit{unbiased}
random walk (RW), bounded by `soft' walls near the extremes of the allowed
region. By contrast, $M\left( t\right) $ for an Ising system below
criticality spends much of its time hovering around the spontaneous
magnetizations, $\pm M_{sp}$, and makes rare and short excursions from one
to the other. Though not shown here, we observe no metastability when the
two nodes `change sides': $\left( 101,99\right) \rightarrow \left(
99,101\right) $. $X/\mathcal{N}$ simply marches from $\thicksim 15\%$ to 
$\thicksim 85\%$ in $\thicksim 3500$ MCS. In other words, on the average, $X$
changes by about two links per MCS. We also considered having $\delta =4$ or 
$6$ `defectors' instead of just two, in systems with $N=400$ and $800$. In
all cases, the average `velocity' is approximately $\delta $ per MCS.
Intuitively, we may attribute this to the action of the $\delta $ extra $E$
nodes, but it remains to be shown analytically. In all respects, there is
absolutely \textit{no barrier} between the two extremes of $X$~!

\begin{figure}[tbp]
\centering
\includegraphics[width=3in]{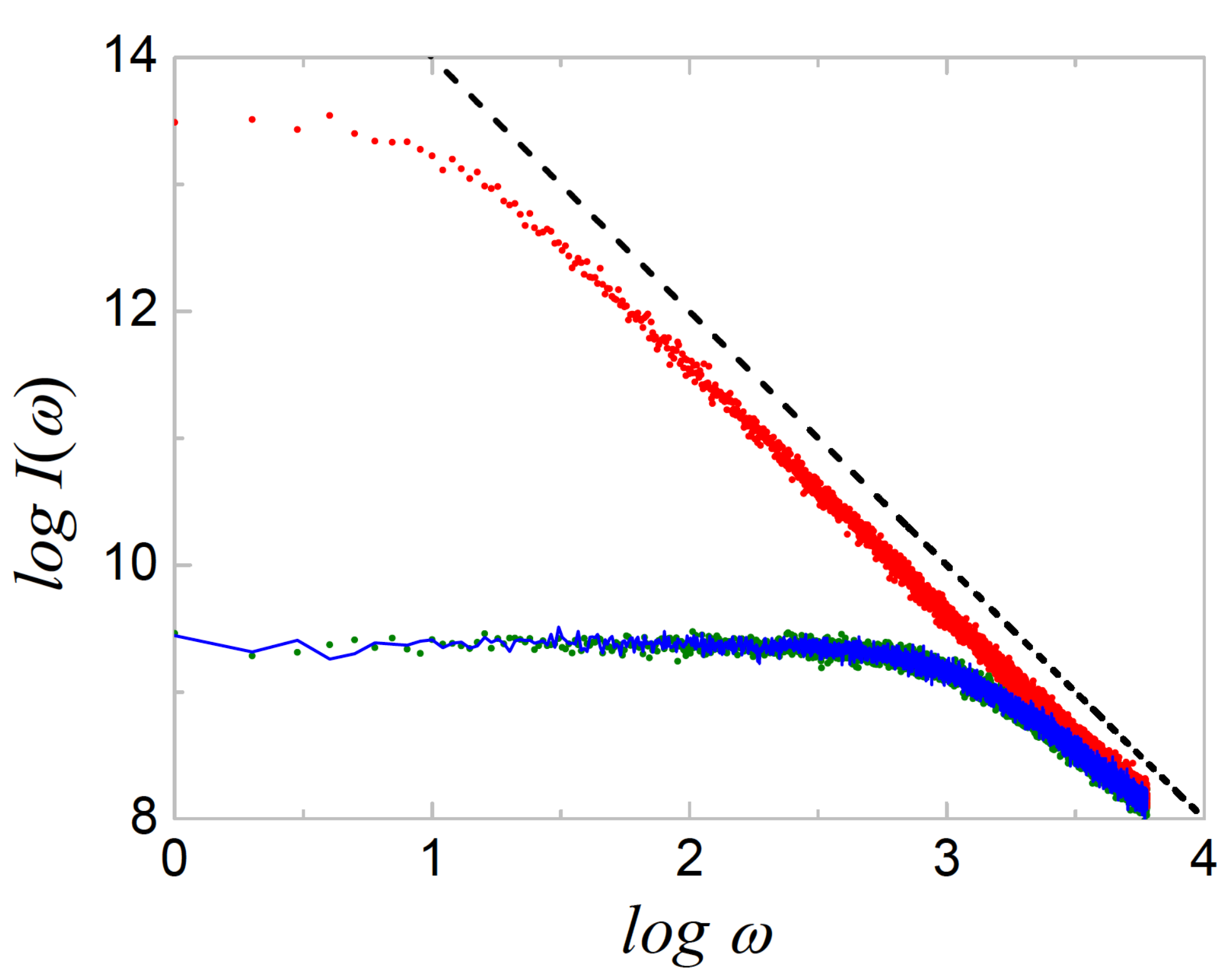}\vspace{-0.5cm}
\caption{ Power spectra, $I\left( \omega \right) $, associated with the time traces in
Fig.~\ref{timetrace}. The dashed black line is proportional to $1/\omega
^{2} $. Note that the spectra associated with the two off-critical cases
(shown as green dots and a blue line) are statistically identical, as
expected from particle-hole symmetry. }
\label{PS}
\end{figure}

Finally, to confirm the notion of a RW, we compute the power spectrum of 
$X\left( t\right) $ as follows. With $\mathcal{T}=2\times 10^{4}$
measurements (of runs of $2\times 10^{6}$ MCS), we compute the Fourier
transform $X\left( \omega \right) $ and then average over 100 runs to obtain 
$I\left( \omega \right) \equiv \left\langle \left\vert X\left( \omega
\right) \right\vert ^{2}\right\rangle $. In Fig.~\ref{PS}, we show plots of 
$\log I$ \textit{vs.} $\log \omega $, as well
a straight line (black dashed) representing $\omega ^{-2}$. The red data
points, associated with $N_{I}=N_{E}=100$, are statistically consistent with
the RW characteristic of $\omega ^{-2}$. The cutoff at small $\omega $ can be
estimated from the finite domain of the RW ($\thicksim 7000$ here). Since 
$\Delta X=\pm 1$ in each attempt, we can assume the traverse time to be 
$\thicksim 7000^2\cong 5\times 10^{7}$ attempts, or 
$\thicksim 2.5\times 10^{5}$ MCS. Given that this value is comparable to 
$1/10$ of our run time, it is reasonable to expect deviations from the pure 
$\omega ^{-2}$ as we approach $\omega \thicksim 10$. 
By contrast, the power spectra of the
two off-critical cases (green dots and blue line, from the green and blue
traces in Fig.~\ref{timetrace}) are controlled by some intrinsic time scale
associated with both the restoration to $\left\langle X\right\rangle $ and
the fluctuations thereabout. Indeed, this $I\left( \omega \right) $ is
entirely consistent with a Lorentzian, i.e., $\propto 1/\left( \omega
^{2}+\omega _{0}^{2}\right) $. Given our limited understanding of the
dynamics of this model, estimating $\omega _{0}$ is beyond the scope of this
work. Let us remark that we have preliminary data for systems of other sizes
and that, while data collapse can be achieved, we have not found simple
explanations for the parameters involved. Instead, we will present a brief
phenomenological scaling analysis of one aspect of the critical system in
Subsection $4.2$. Here, let us turn to a mean-field approach which
can provide some understanding of a number of these unusual properties.

\subsubsection{Mean-field approximation}

Unlike the complications we face with the degree distributions, it is simple
to formulate a mean-field theory for $X$, directly from $\mathcal{P}^{ss}$.
The underlying difference is that there is only one variable here and, in
the spirit of the mean-field approximation, we can simply replace every $n_{ij}$
by $\left\langle X\right\rangle /\mathcal{N}$. Indeed, we may study the full
(steady state) distribution of $X$, given exactly by%
\begin{equation}
P\left( X\right) \equiv \sum_{\left\{ \mathbb{N}\right\} }\delta \left(
X,\Sigma _{ij}n_{ij}\right) \mathcal{P}^{ss}\left( \mathbb{N}\right) .
\label{PX}
\end{equation}%
Performing this sum to a closed form, however, is not feasible so far, since
this task is comparable to finding $P\left( M\right) $ for the 2-dimensional
Ising model. Therefore, we attempt a mean-field approach to make progress,
replacing $k_{i}=\Sigma _{j}n_{ij}$ by $N_{E}\left( X/\mathcal{N}\right)
=X/N_{I}$ and $p_{j}=\Sigma _{i}\bar{n}_{ij}$ by $N_{I}\left( 1-X/\mathcal{N}%
\right) =N_{I}-X/N_{E}$. Meanwhile%
\begin{equation}
\sum_{\left\{ \mathbb{N}\right\} }\delta \left( X,\Sigma _{ij}n_{ij}\right) =%
\binom{\mathcal{N}}{X}  \label{entropy}
\end{equation}%
so that our approximate distribution, $\tilde{P}$, is%
\begin{equation}
\tilde{P}\left( X\right) \propto \binom{\mathcal{N}}{X}\left( \left[ \frac{X%
}{N_{I}}\right] !\right) ^{N_{I}}\left( \left[ N_{I}-\frac{X}{N_{E}}\right]
!\right) ^{N_{E}}
\end{equation}%
In this spirit, it is natural to consider%
\begin{equation}
F\left( x;N_{I},N_{E}\right) \equiv -\ln \tilde{P}\left( X\right) /\mathcal{N%
}  \label{Landau FE}
\end{equation}%
and regard it as a `Landau free energy density' for the intensive variable 
\begin{equation}
x\equiv X/\mathcal{N}\in \lbrack 0,1]
\end{equation}

In the thermodynamic limit, the leading order of $F$ is \textit{linear} in 
$x $, with slope $\ln (N_{I}/N_{E})$ \cite{CSP25,LiuSchmittmannZia12}. Thus,
as long as $N_{I}\neq N_{E}$, $x$ can only assume boundary values, $0$ or $1$,
while, for $N_{I}=N_{E}$, $F$ is \textit{flat} over the entire interval.
This prediction agrees qualitatively with the main simulation results,
especially the evolution $\Delta X/\Delta t\thicksim \delta $/MCS when 
$N_{E}=N_{I}-\delta $ is suddenly changed to $N_{E}=N_{I}+\delta $. Keeping
the next order, 
$-\left. \left( \frac{\ln x}{N_{E}}+\frac{\ln (1-x)}{N_{I}}\right) \right/ 2$,
we find not only nonlinear terms, but also the necessary
`repulsion' to avoid the extremes. At this order, $x$ settles within 
$O\left( 1/N\ln \left[ N_{I}/N_{E}\right] \right) $ of the boundaries for
generic $N_{I,E}$. Cast in the language of magnetism, $F\left( m;h,N\right) $
reads%
\begin{equation*}
\frac{m}{2}\ln \frac{1+h}{1-h}-\frac{1}{N}\left[ \frac{\ln \left( 1+m\right) 
}{1+h}+\frac{\ln \left( 1-m\right) }{1-h}\right] +...
\end{equation*}%
From here, we can find the minimum of $F$ and obtain a mean-field `equation
of state':%
\begin{equation}
m\left( h\right) \cong -H+sign\left( h\right) \sqrt{H^{2}-2Hh+1}
\label{m(h)}
\end{equation}%
where 
\begin{equation}
H^{-1}=\frac{1}{2}N\left( 1-h^{2}\right) \ln \frac{1+h}{1-h}=\frac{%
2N_{E}N_{I}}{N}\ln \frac{N_{E}}{N_{I}}
\end{equation}%
may be regarded as an alternate definition of a `magnetic field' like
control variable.

How does this prediction compare to data? For the specific case of $N=200$,
this $m\left( h\right) $ (solid black curve in Fig.~\ref{mh}) is remarkably
close to the data points. Clearly, this mean-field approach captures some
key features of the $XIE$ model. We should caution the reader, however, that
this plot shows a deceptively good agreement. Indeed, this prediction
deviates from data (e.g., by as much as $3\%$ for the $\left( 110,90\right) $
case) considerably more than those from the self-consistent mean-field
theory (where $\left\langle X\right\rangle =\left\langle k_{I}\right\rangle
N_{I}$). Nor does the good agreement extend to the entire distribution: $%
\tilde{P}\left( X\right) $ deviates substantially from the histograms of $X$%
, especially for $N_{E}\cong N_{I}$. Nevertheless, it does offer some
insight into the major differences between this system and an Ising
ferromagnet at low temperatures. Note that, if the $N\rightarrow \infty $
limit is taken first in this approach, we find a highly singular $m\left(
h\right) =sign\left( h\right) $. Work is in progress to explore the how such
an extreme equation of state can be understood quantitatively in the context
of an extreme Thouless effect \cite{BarMukamel14}.

\subsection{A scaling study of the `critical' system: $N_{I}=N_{E}$}

In this subsection, we report preliminary results for a scaling study of the
`critical' system, $N_{I}=N_{E}$. For convenience, let us define $L\equiv
N_{I,E}=N/2$, so that we can regard our system as an $L\times L$ Ising
model with $\mathcal{N}=L^{2}$ `spins' while $M=2X-L^{2}$ is its total
magnetization. As can be seen in Fig.~\ref{histogram}, the
distribution $P(X)$ displays a broad plateau. In terms of the fraction 
$X/\mathcal{N}$, the width of the plateau increases with $L$. We perform 
extensive simulations to investigate this phenomenon quantitatively. 
Specifically, we determine the location of the left edge of the plateau 
as a function of $L$, defined as the value of $X$ ($<\mathcal{N}/2$) where 
$P$ has vanishing curvature. Thus, let us define $X_{Q_{\mathrm{max}}}(L)$
as the value where%
\begin{equation*}
Q(X)=\frac{dP(X)}{dX}
\end{equation*}%
reaches a maximum value. Determining $X_{Q_{\mathrm{max}}}$ from simulations
accurately poses two challenges, especially for large $L$. First, in the
critical state, $X$ performs a RW over the plateau, spending little time
near the edges. Thus obtaining good statistics for $P(X)$ near the edges is
difficult. Second, taking numerical derivatives of noisy data to obtain an
accurate estimate of $Q(X)$ is non-trivial. We have utilized specialized
numerical methods in an effort to address both of these difficulties.

\begin{figure}[tbp]
\centering
\includegraphics[width=3.5in]{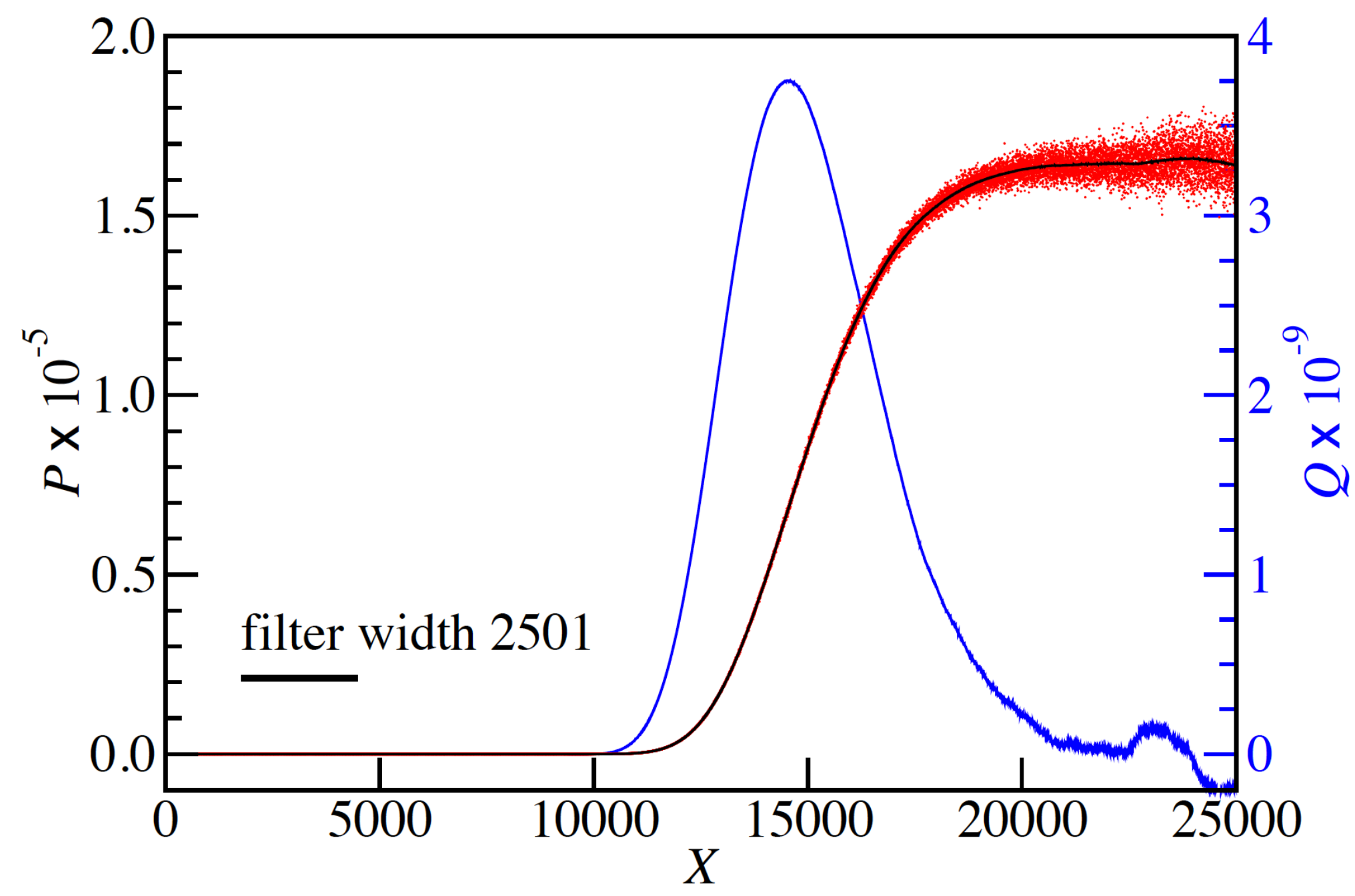}\vspace{-0.4cm}
\caption{Determining the location of the edge of the plateau of the
distribution $P(X)$. An example of the use of Savitzky-Golay filtering to
find $P(X)$ (black dashed line) and its derivative $Q(X)$ (blue solid line)
from data for system size $N=316$ shown as the (red) scattered points. A
filter width slightly less than the full width at half maximum of the peak
of $Q(X)$ was used.}
\label{smoothing}
\end{figure}

In order to alleviate the first difficulty, we exploited the fact that the
original dynamics sends our system to an equivalent equilibrium system with
a Hamiltonian given by Eqn.~(\ref{H}) (and unit temperature in the Boltzmann
factor). Thus, we are allowed to bias the dynamics by adding an additional
term%
\begin{equation*}
\Delta \mathcal{H}(X)=\mu \left\vert X-\mathcal{N}/2\right\vert
\end{equation*}%
to $\mathcal{H}$. With $\mu >0$, this extra term biases the system towards
the edges of the plateau in $P(X)$, against spending time in the center. As
a result, in a simulation of the modified system, the region near the edges
is sampled much more frequently, providing much better statistics. To
recover the desired $P\left( X\right) $, from $P_{\Delta \mathcal{H}}(X)$,
the distribution of the modified system requires a simple reweighting \cite
{FerSwe88}:
\begin{equation*}
P(X)\propto P_{\Delta \mathcal{H}}(X)\;e^{-\Delta \mathcal{H}(X)}.
\end{equation*}

Even with such improvements, the data for $P(X)$ is still quite noisy for
the purpose of obtaining its derivative $Q$ reliably. One possibility is to
simply bin the data for $P(X)$, but this procedure could shift the location
of $X_{Q_{\mathrm{max}}}$. To overcome this second difficulty, we smooth the
data using \textit{Savitzky-Golay filtering} (SGf)~\cite{SavGol64,NumRec}.
This method of filtering performs a least-squares fit of a polynomial
function over a moving window, the `filter width', of the
data. A particularly useful feature of SGf, as opposed to other related
filtering methods, is that it efficiently produces a smoothed derivative
function that fits the data as well as the function itself. An example of
our use of SGf is shown in Fig.~\ref{smoothing}. In the figure, we show the
reweighted data for $P(X)$ (for $L=316$), as well as the results for the
fitted functions $P(X)$ and $Q(X)$. We explore using polynomials of
different order and using different filter widths in our smoothing process.
For the sigmoidal shaped data we have, we find stable results for locating 
$X_{Q_{\mathrm{max}}}$ by using a fourth-order polynomial and a filter width
that is slightly less than the full width at half maximum of the peak in 
$Q(X)$.

\begin{figure}[tbp]
\centering
\includegraphics[width=3.5in]{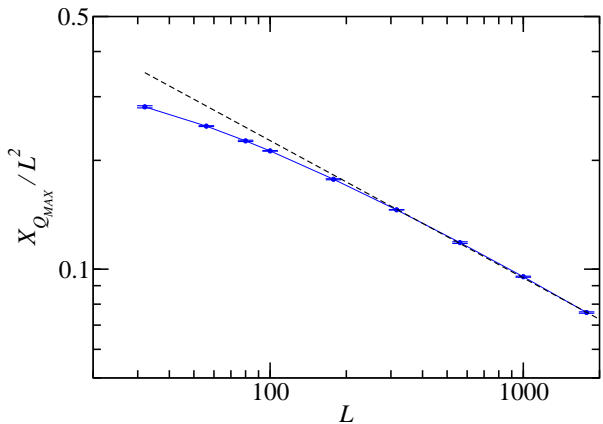}\vspace{-0.8cm}
\caption{Scaling behavior of $X_{Q_{\mathrm{max}}}$: the location of the edge
of the plateau of the distribution $P(X)$ as a function of system size $N$. }
\label{wallscaling}
\end{figure}

Using these methods, we determine $X_{Q_{\mathrm{max}}}(L)$ for $L$ ranging
from 32 to 1778. For each system size, we carry out three to five
different simulation runs, ranging from $10^{11}$ MCS for the small systems
to $5\times 10^{9}$ MCS for the largest. The values of $\mu $ used in the
modified simulations and subsequent reweighting vary from about $10^{-3}$
for $L=100$ (modified simulations are not used on smaller systems) to about 
$10^{-6}$ for $L=1778$. Each run is analyzed separately, so that we have an
estimate of the statistical errors. The results are shown as a log-log plot
of $X_{Q_{\mathrm{max}}}/L^{2}$ vs. $L$ in Fig.~\ref{wallscaling}. The error
bars indicate the $2\sigma $ statistical errors in the variation of the
results from the different runs. Systematic errors associated with changing
the filtering parameters are in the same range. The straight line in the
figure has a slope of $-0.38$. The conclusion is that, for large systems,
the \textit{fraction} of cross-links at the left edge of $P\left( X\right) $
scales as a power of $L:$%
\begin{equation*}
\frac{X_{Q_{\mathrm{max}}}}{L^{2}}\sim L^{-0.38}.
\end{equation*}

Of course, it is difficult to compare such a result to those from the 
mean-field theory above. Clearly, despite its minimal nature, this system
displays a variety of rich behavior, understanding which will require, at
the least, a detailed finite size analysis.

\section{Summary and outlook}

In this article, we continue our study of preferred degree networks and
the interaction between such networks, but focus on a particular limit,
with \textit{extreme} introverts and extroverts (i.e., $\kappa _{I}=0$ and $%
\kappa _{E}=\infty $, respectively). In this minimal model, there are only
two control parameters, $N_{I,E}$, the number of $I$'s and $E$'s. Since they
only cut or add connections if possible, the intra-community links quickly
empty/fill and become frozen. As only the cross-links are active, each
configuration of the network can be completely specified by an incidence
matrix, $\mathbb{N}$. ($n_{ij}=1,0$ for the presence or absence of a link
between nodes $i$ and $j$). Though the dynamics of interacting networks of
this class does not generally obey detailed balance, it is restored in this
limit, so that we are able to find an analytic expression for the
stationary, microscopic distribution, $\mathcal{P}\left( \mathbb{N}\right) $.
Since the elements of $\mathbb{N}$ are binary, our system can be regarded
as a kinetic Ising model and $-\ln \mathcal{P}$ can be thought of as a
`Hamiltonian' $\mathcal{H}$. In this language, our system displays peculiar
long range, `multi-spin' interactions, unlike any magnetic systems in nature.
Like the Ising model, a particle-hole symmetry prevails here, though it is
slightly more obscure.

Our simulation study of its collective behavior focuses on the degree
distributions and $X$, the total number of cross-links, relying mostly on
systems with $N_{I}+N_{E}=200$. The most remarkable aspect is that, even
when $N_{I}\cong N_{E}$, the fraction of cross-links is far from the
symmetric value of $1/2$. Instead, there is a sizeable jump as $N_{I}$
crosses $N_{E}$, e.g., from $\thicksim 15\%$ to $\thicksim 85\%$ when
just two nodes `change sides' ($\left( 101,99\right) \rightarrow \left(
99,101\right) $). The nature of this transition fails to conform to the
standard categories of first or second order. On one hand, the sizeable jump
of $\langle X\rangle $ suggests a first order phase transition. Yet, the
other features typically associated with first order phase transitions are
absent here, such as metastability, hysteresis, phase co-existence, etc. On
the other hand, the extensive fluctuations and slow dynamics are more
typical of a system at a second order transition. Most likely, it belongs to
the class of less well-known, mixed-order transitions.
Indeed, it is very likely that the 
$XIE$ displays an `extreme' Thouless effect \cite{BarMukamel14}, in that the
jump of $\left\langle X\right\rangle /\mathcal{N}$\ will cover the entire
unit interval as $N\rightarrow \infty $. We expect such behavior from the
limited finite size analysis presented here, namely, the left edge of the
plateaux in $P\left( X\right) $ for the symmetric case decreases as 
$1/N^{0.38}$. Work towards a more systematic study of finite size scaling is
in progress \cite{BasslerZia14}. Naturally, these extraordinary aspects are
expected to manifest in the more detailed degree distributions. So far,
apart from the $N_{I}=N_{E}$ case, the degree distributions can be
reasonably well predicted by a self-consistent mean-field approach. A less
refined, but more accessible, approximation scheme is presented for
describing the behavior of $X$. Though less accurate, this scheme offers
some insight into the puzzling features associated with the transition.
There is clearly room for quantitative
improvements. 
Naturally, these extraordinary aspects are also manifest in the more
detailed degree distributions. Apart from the $N_{I}=N_{E}$ case, the degree
distributions can be reasonably well predicted by a self-consistent mean
field approach. A less refined, but more accessible, approximation scheme is
presented for describing the behavior of $X$. Though less accurate, this
scheme offers some insight into the puzzling features associated with the
transition.

This remarkably simple, yet quite rich, system deserves to be investigated
further. The most obvious question may be whether a thermodynamic limit
exists, and in what way. For example, do the degree distributions approach a
non-trivial limit when $N\rightarrow \infty $? with $N_{I}/N_{E}$ held fixed
or with $N_{I}-N_{E}$ being a constant? as a function of $k,p$ or some
scaled variable? Despite the presence of an explicit expression for the
microscopic distribution and its associated $\mathcal{H}$, computing the
partition function, let alone averages of quantities of interest, poses a
worthy challenge. The failure of mean-field theory, especially near
`criticality' hints at the importance of correlations. Preliminary studies
indicate strong correlations, in the sense that the difference between the
joint distribution, $\rho \left( k_{I},k_{E}\right) $, and the product $\rho
_{I}\left( k_{I}\right) \rho _{E}\left( k_{E}\right) $ can be a sizeable
fraction of the former. Systematic investigations of them are straightforward
and surely worthwhile. Presumably, once some progress along these directions
is made, it is possible to investigate the more serious issues concerning
the nature of the mixed-order transition. For example, can we understand and
predict the absence of \textit{any}\textbf{\ }barrier (in the symmetric
system) between the degenerate extremes? Launching a systematic study of
finite size scaling will also help to shed light on such peculiarities of
this model. Of course, we should also pursue the questions raised above,
e.g., details of the power spectra, especially for systems with $N_{I}\neq
N_{E}$. In an earlier paper of this series \cite{LiuSchZia14}, we introduced
a quantitative concept of `frustration' (in the ordinary psychological
sense). Though $XIE$ seems to be a maximally frustrated model, its role and
behavior are yet to be considered.

Beyond exploring these questions, we can extend the $XIE$ mode in an
orthogonal direction, arguably of purely theoretical interest (at present).
We may treat $\mathcal{H}$ as a genuine Hamiltonian in a standard study of
critical phenomena in thermal equilibrium. In other words, we propose to 
study the statistical mechanics of a $L\times L$ system associated with 
the Boltzmann factor%
\begin{equation}
\mathcal{P}\propto \exp \left\{ -\beta \left[ \mathcal{H}-BX\right] \right\}
\end{equation}%
where $B$ plays the role of a symmetry breaking, `magnetic field' (as
opposed to $h$ in the $XIE$). It is interesting to note that, while the
critical control parameters of a typical system (e.g., $T_{c}$ in Ising, $%
T_{c}$ and $P_{c}$ for liquid gas) are not known, they are given precisely
by $\beta _{c}=1$ and $B_{c}=0$ here. For this `purely theoretical' system,
work is in progress \cite{BasslerZia14} to explore the usual avenues of 
interest: static and dynamic critical exponents, scaling functions, 
universality and the classes, etc. In the context of renormalization group 
analyses (which proved to be highly effective in dealing with other 
mixed-order transitions \cite{BarMukamel14} ), we already know that 
$\mathcal{H}$ lies on the critical sheet and can
inquire about fixed points and their neighborhoods, irrelevant and relevant
variables (e.g., if there are others besides $\beta -1$ and $B$). Similarly,
we might ask if the perspectives from catastrophe theory \cite{Thom} can offer
fresh insight. 

Beyond the $XIE$ and its companion model, there is a wide vista involving
dynamic networks with preferred degrees. For instance, instead of assigning one
or two $\kappa $'s to a population, it is surely more natural to assign a
distribution of $\kappa $'s. There are also multiple ways to model
interactions between the various groups. For example, even with just two
groups, it is realistic to believe that an individual may have \textit{two }%
preferred degrees, one for contacts within the group and another for those
outside. Surely, this kind of differential preference underlies the formation
of social cliques. Beyond understanding the topology and dynamics of
interacting networks of the types described here, the next natural step is
to take into account the freedom associated with the nodes, e.g., opinion,
wealth, health, etc., on the way to the ambitious goal of understanding
adaptive, co-evolving, interdependent networks in general.

\appendix

\section{Restoration of detailed balance}

In this appendix, we show that all Kolmogorov loops are reversible in the
XIE model and so, detailed balance is restored \cite{kolmogorov}. 
Since the full dynamics occurs on the $\mathcal{N}$ cross-links, 
the configuration space consists of
the corners of an $\mathcal{N}$-dimensional unit cube, while adding/cutting
a link is associated to traverse along an edge therein. Clearly, products of
the ratios of forward and reversed transition rates around any closed loop
can be expressed in terms of those around `elementary loops' -- i.e., loops
around a plaquette on the $\mathcal{N}$-cube. We will show that the ratio
associated with every plaquette is unity and so, all Kolmogorov loops are
reversible.

First, it is easy to see that if an elementary loop consists of modifying
two links connected to four different nodes, then the actions on each link
are unaffected by the other. In other words, rates associated with opposite
sides of the square (loop) are the same. Thus, their product in one
direction is necessarily the same as in the reverse. We need to focus only
on situations where the two links are connected to three nodes, e.g., $ij$
and $im$. For any such loop, let us start with a configuration in which both
are absent ($n_{ij}=n_{im}=0$). Let the states of node be such that $i$ has $%
k_{i}$ links, and $j$ and $m$ have $p_{j}$ and $p_{m}$ `holes',
respectively. Then one way around the loop is adding these two links
followed by cutting them, which can be denoted as the sequence 
\begin{equation}
\binom{n_{ij}}{n_{im}}=\binom{0}{0}\rightarrow \binom{1}{0}\rightarrow 
\binom{1}{1}\rightarrow \binom{0}{1}\rightarrow \binom{0}{0}
\end{equation}%
and leaving the rest of $\mathbb{N}$ unchanged. The associated product of
the transition rates is, apart from an overall factor of $N^{4}$, 
\begin{equation}
\frac{1}{p_{j}}\frac{1}{p_{m}}\frac{1}{k_{i}+2}\frac{1}{k_{i}+1}
\label{product}
\end{equation}%
Now, the reversed loop can be denoted as 
\begin{equation}
\binom{n_{ij}}{n_{im}}=\binom{0}{0}\rightarrow \binom{0}{1}\rightarrow 
\binom{1}{1}\rightarrow \binom{1}{0}\rightarrow \binom{0}{0}
\end{equation}%
associated with the product 
\begin{equation}
\frac{1}{p_{m}}\frac{1}{p_{j}}\frac{1}{k_{i}+2}\frac{1}{k_{i}+1}
\end{equation}%
which is exactly equal to Eqn.~(\ref{product}). From symmetry, we can expect the
same results for loops involving two introverts and one extrovert (i.e., $ij$
and $kj$). Thus, we conclude that the Kolmogorov criterion is satisfied and
detailed balance is restored in this XIE limit. Our system should settle
into a stationary distribution without probability currents, much like the
Boltzmann distribution for a system in thermal equilibrium.

\begin{acknowledgments}
We thank D. Dhar, Y. Kafri, W. Kob, D. Mukamel, Z. Toroczkai for illuminating discussions. This research is supported in part by the US National Science Foundation, through grants DMR-1206839 and DMR-1244666, and (KEB) by AFOSR and DARPA through grant FA9550-12-1-0405. One of us (RKPZ) thanks the Galileo Galilei Institute for Theoretical Physics for hospitality and the INFI for partial support during the completion of this paper.
\end{acknowledgments}


\end{document}